\documentclass[preprint,showpacs,preprintnumbers,amsmath,amssymb]{revtex4}
\usepackage{graphicx}
\usepackage{tikz}
\usepackage{amsmath}
\usepackage{amsfonts}
\usepackage{url}

\def\bxA{{\mathbf{x}}_A}
\def\bxS{{\mathbf{x}}_S}

\def\s0{{\mathbb{S}_0}}

\def\d3{{\partial_3}}

\newcommand{\eqnref}[1]{Equation \eqref{#1}}
\newcommand{\figref}[1]{Figure \ref{#1}}

\graphicspath{{/home/joeri/Documents/Python/2D/resamp/imag/}}

\begin{document}
\textbf{Data-driven internal multiple elimination applications using imperfectly sampled reflection data}\\
\\
\textbf{Joeri Brackenhoff, Johno van IJsseldijk and Kees Wapenaar}\\
\\
\textbf{Abstract.} We consider reflection data that have been subsampled by 70\% and use Point-Spread-Functions to reconstruct the original data. The subsampled, original and reconstructed reflection data are used to image the medium of interest with the Marchenko method. The image obtained using the subsampled data shows artifacts caused by internal multiples, which are eliminated when the original and reconstructed data are used.
\newpage
\section{Introduction}
\vspace{-0.3cm}
In recent years, the data-driven Marchenko method has frequently been used to remove internal multiples. The method employs reflection data, measured at an open boundary at the surface of the Earth, and an estimation of the first arrival from a location of interest in the subsurface \citep{wapenaar2014marchenko}. While the method has shown robust results, the requirements for the reflection data are demanding. \cite{staring2018source} and \cite{Brackenhoff2019virtual} showed that in order to obtain good results for the purpose of imaging and homogeneous Green's function retrieval, respectively, the data need to be densely sampled in a spatial sense. In case the reflection data have poor spatial sampling, the results of the applications degrade. Recently, \cite{wapenaar2019discrete} showed that by using Point-Spread-Functions (PSFs), the imperfect sampling of the sources of the reflection data can be accounted for. The authors showed that using this approach, the Green's function can be reconstructed with less artifacts caused by the poor spatial sampling. In this paper, we apply this approach to reconstruct the sources of sparsely sampled reflection data, in order to use it for various Marchenko-based applications. We demonstrate this approach by imaging, using reconstructed 2D synthetic reflection data and comparing the result to the one obtained when sparsely sampled data are used, as well as when a densely sampled dataset is used.
\vspace{-0.3cm}
\section{Imaging using reconstructed reflection data}
\begin{figure}[htpb]
\vspace{-0.4cm}
\centering
\begin{tikzpicture}
\node at (0,0) {\includegraphics[width=1.0\columnwidth]{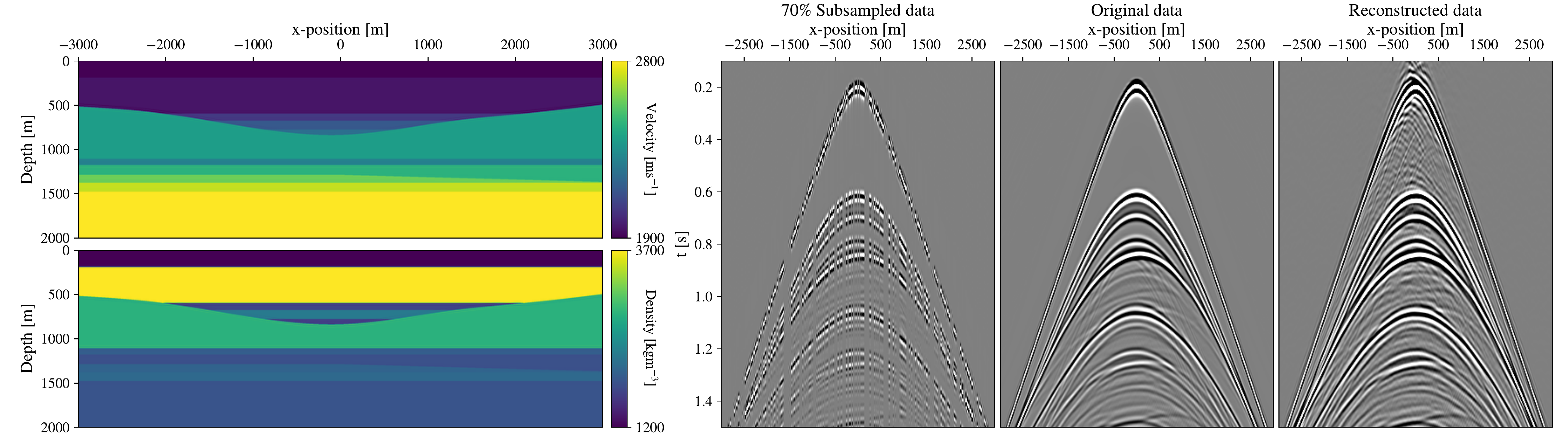}};
\node[fill=white,draw=black,inner sep=1pt] at (-7.15,1.45) {\tiny (a)};
\node[fill=white,draw=black,inner sep=1pt] at (-7.15,-0.5) {\tiny (b)};
\node[fill=white,draw=black,inner sep=1pt] at (-0.40,1.45) {\tiny (c)};
\node[fill=white,draw=black,inner sep=1pt] at (-0.40+2.93,1.45) {\tiny (d)};
\node[fill=white,draw=black,inner sep=1pt] at (-0.40+2.93*2,1.45) {\tiny (e)};
\end{tikzpicture}
\vspace{-1.0cm}
\caption{\textbf{(a)} Velocity in $ms^{-1}$ and \textbf{(b)} density in $kgm^{-3}$ of the medium of interest. Common-receiver record, located at the center of the model, of \textbf{(c)} the subsampled reflection data, \textbf{(d)} the original reflection data and \textbf{(e)} the reconstructed reflection data. All reflection data have been convolved with a 25 Hz Ricker wavelet.}
\label{refl}
\end{figure}
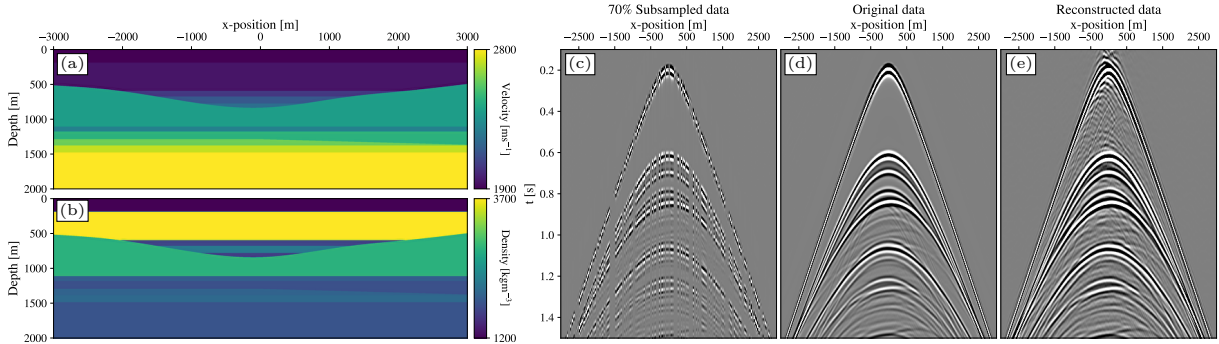

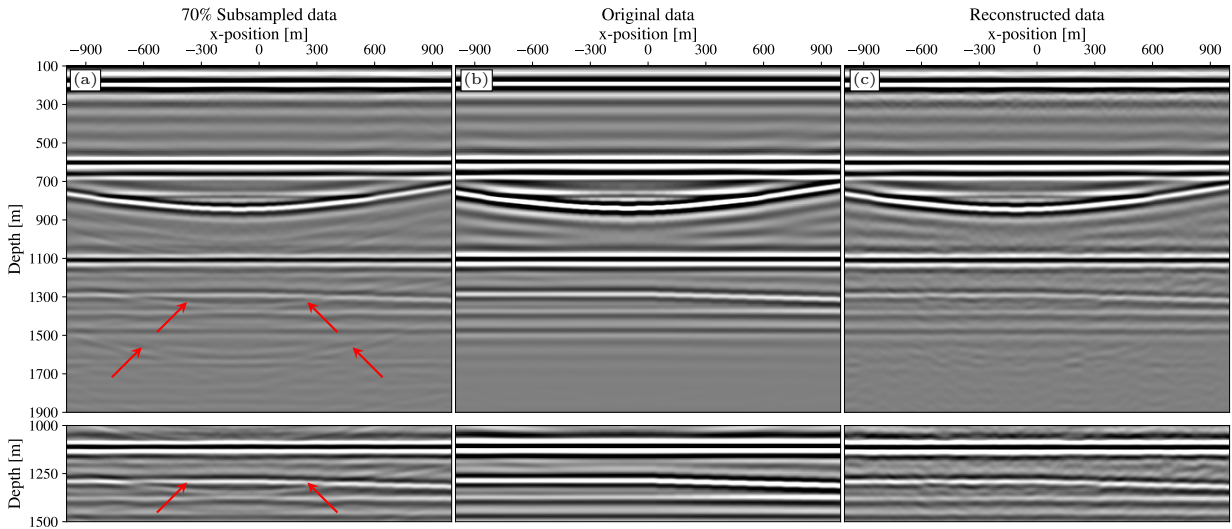
\begin{figure}[htpb]
\vspace{-0.5cm}
\centering
\begin{tikzpicture}
\node at (0,0) {\includegraphics[width=1.0\columnwidth]{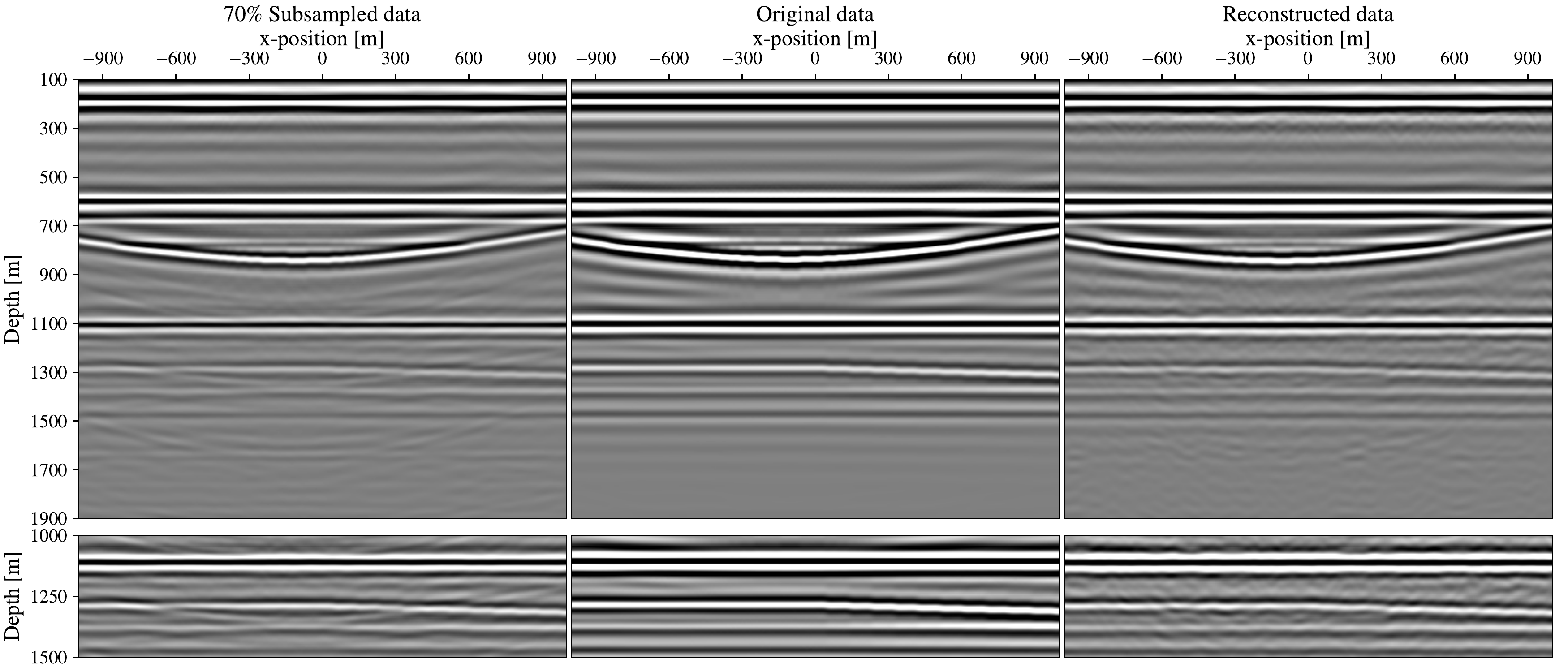}};
\node[fill=white,draw=black,inner sep=1pt] at (-7.15,2.5) {\tiny (a)};
\node[fill=white,draw=black,inner sep=1pt] at (-7.15+5.2,2.5) {\tiny (b)};
\node[fill=white,draw=black,inner sep=1pt] at (-7.15+5.18*2,2.5) {\tiny (c)};
\draw[red,thick,>=stealth,->] (-6.2,-0.85) -- ++(0.4,0.4);
\draw[red,thick,>=stealth,->] (-6.8,-1.45) -- ++(0.4,0.4);
\draw[red,thick,>=stealth,->] (-6.2,-3.25) -- ++(0.4,0.4);
\draw[red,thick,>=stealth,->] (-3.8,-0.85) -- ++(-0.4,0.4);
\draw[red,thick,>=stealth,->] (-3.2,-1.45) -- ++(-0.4,0.4);
\draw[red,thick,>=stealth,->] (-3.8,-3.25) -- ++(-0.4,0.4);
\end{tikzpicture}
\vspace{-1.1cm}
\caption{Image of the subsurface obtained using \textbf{(a)} the subsampled reflection data, \textbf{(b)} the original reflection data and \textbf{(c)} the reconstructed reflection data. Below each panel a deep section of the image is shown, with a clipping factor that is 2 times stronger. All data have been convolved with a 25 Hz Ricker wavelet. The red arrows in \textbf{(a)} indicate false reflections.}
\label{imag}
\end{figure}

\vspace{-0.3cm}
The velocity and density model for the numerical experiment are shown in \figref{refl} \textbf{(a)} and \textbf{(b)}, respectively. The original reflection data are computed, using densely sampled receivers and sources (\figref{refl}\textbf{(d)}). Next, these data are subsampled by removing 70\% of the receivers at random (\figref{refl}\textbf{(c)}). Finally, the data need to be reconstructed. To achieve this, a virtual datum at 100m depth is defined. The blurred version of the upgoing Green's function is retrieved at this datum from the subsampled reflection data with a single iteration of the Marchenko scheme. Following \cite{wapenaar2019discrete}, this Green's function can be deblurred using Point-Spread Functions (PSFs). Subsequently, the reconstructed reflection data ($\hat{R}$) are computed from the direct arrival of the Green's function ($G_d^{+,+}$) and the PSF-corrected upgoing Green's function ($\hat{G}^{-,+}$) due to a downgoing source, as follows \citep{vanderneut2016adaptive}:
\begin{equation} \label{rec}
\hat{R}({\mathbf{x}}_S,{\mathbf{x}}_R,t)=\int_{\partial\mathbb{D}_A}G_d^{+,+}(\bxA,{\mathbf{x}}_S,t)\ast \hat{G}^{-,+}(\bxA,{\mathbf{x}}_R,t){\rm d^2}\bxA,
\end{equation}
In this equation $t$ is the time, ${\mathbf{x}}_S$ and ${\mathbf{x}}_R$ denote the source and receiver positions on the Earth's surface, respectively, $\bxA$ represents the virtual locations inside the subsurface (on $\partial\mathbb{D}_A$) and $\ast$ indicates a convolution. The reconstructed reflection data are shown in \textbf{(e)}.\\
To demonstrate the effectiveness of our approach, we use the reconstructed reflection data to image the model of interest. We image the model using the double-focusing method \citep[eq. (11)]{staring2018source}:
\begin{equation} \label{df}
G^{-,+}(\bxA,\bxA',t)=\int_{\partial\mathbb{D}_0}G^{-,+}(\bxA,\bxS,t)\ast f_1^+(\bxS,\bxA',t){\rm d^2}\bxS,
\end{equation}
where $f_1^+$ is the downgoing focusing function of the first type and $\partial\mathbb{D}_0$ is the boundary at the surface of the Earth. \eqnref{df} states that if one has a focusing function and decomposed Green's function, with a redatumed receiver, obtained through the focusing of the Marchenko method, the source of the decomposed Green's function can be redatumed. By coinciding the source location ($\bxA'$) and receiver location ($\bxA$) of the decomposed Green's function at a location inside the medium of interest, the local reflectivity can be obtained \citep{staring2018source}. The required Green's function and focusing function for \eqnref{df} can be obtained using the Marchenko method, and we use a similar approach as \cite{Brackenhoff2019virtual} to achieve this, by smoothing the velocity model and using an eikonal solver.\\
We use \eqnref{df} and the three sets of reflection data to create images of the subsurface. The results are shown in \figref{imag}. The result in \textbf{(b)}, using the original reflection data shows all layer contrasts clearly without internal multiples. If the subsampled reflection data are used, the internal multiples are not properly eliminated, as can be seen from the false reflections indicated by the red arrows in \textbf{(a)}. If we use the reconstructed reflection data, the result is much closer to that using the original data, as shown in \textbf{(c)}. The differences become clearer, when considering a deep part of the model with a potential reservoir, shown in panels below the full images, and setting a stronger clipping factor. While the reconstructing does account for most of the multiple energy, some artifacts remain. Although these artifacts are undesired, the reconstructing of the reflection data causes an overall improvement of the imaging result.
\vspace{-0.3cm}
\section{Conclusion}
\vspace{-0.3cm}
We have shown that by reconstructing the reflection data at the surface of the Earth using PSFs, the application of the Marchenko method can be improved. We have shown this by removing 70\% of the samples of a synthetic 2D dataset and comparing the result before reconstructing, after reconstructing and by using the original data. The reconstructing of the reflection data helps to eliminate artifacts caused by internal multiples.
\\%
\\%
\textbf{Acknowledgements.} This research was funded by the European Research Council (ERC) under the
European Union's Horizon 2020 research and innovation programme (grant agreement No: 742703).
{\footnotesize\bibliography{eage}}
\end{document}